\documentclass[a4paper,11pt]{article}
\pdfoutput=1 % if your are submitting a pdflatex (i.e. if you have
\usepackage{jinstpub} % for details on the use of the package, please
\title{\boldmath Interpretation of coincidence data from strip detectors and extraction of absolute cross section using 3-body Monte Carlo simulation method}

\author[a,b]{S.~K.~Pandit\note{Corresponding author.}}
\author[a,b]{K.~Mahata}
\author[a,b]{A.~Shrivastava}
\author[a]{V.~V.~Parkar}
\author[a]{K.~Ramachandran}
\affiliation[a]{Nuclear Physics Division, Bhabha Atomic Research Centre, Mumbai - 400085, India}
\affiliation[b]{Homi Bhabha National Institute, Anushaktinagar, Mumbai - 400094, India}
\emailAdd{sanat@barc.gov.in}

\abstract{Accurate knowledge of the response of the detection system is very crucial for unambiguous interpretation of the experimental data. A simulation code has been developed using the Monte Carlo technique involving 3-body kinematics for the analysis of data obtained with segmented large area Si $\Delta E-E$ detector telescopes in nuclear reaction measurements. Care was taken in the analysis to maximize the angular coverage and statistics. The emphasis is placed to extract the absolute cross sections of the different reaction processes, for which coincident measurements are unavoidable. The estimated detection efficiency of different coincidence events are found to depend on various parameters, e.g., the relative energy of the breakup fragments, incident beam energy of the projectile, ground state $Q$-value of the reaction, the excitation of the ejectile as well as target like nuclei, mass asymmetry of the breakup fragments, detection threshold, and geometric solid angle of the detection setup. The interpretation of the various observables from the exclusive measurements of breakup and transfer breakup reactions is reported.}


\begin{document}
\maketitle
\flushbottom

\section{Introduction}\label{intro}
The study of reactions with radioactive nuclei, which exhibits exotic features like extended matter distribution, halo and Borromean structure, low lying continuum, etc. is a topic of current interest~\cite{Kola16,Keel09,Keel07,Tani85PRL,Auma12}. To study these nuclei special care in measurement as well as theoretical techniques are needed. Currently, there are only few radioactive ion beam (RIB) facilities in use and the beam current is also very low. Whereas, the weakly bound stable nuclei $^{6,7}$Li and $^9$Be have low breakup threshold (1.47 - 2.47 MeV) and available with high intensity and good beam quality. Hence, the investigations involving weakly bound stable nuclei provide the platform to study the influence of breakup/ low lying continuum in reaction dynamics with better statistics. The experimental as well as theoretical knowledge gained from such studies will be useful to study the role of the other exotic features of the radioactive nuclei. The understanding of the reaction dynamics of weakly bound stable nuclei has also an implication on the astrophysical interest~\cite{Kola16,Keel09,Keel07}. 

One of the important aspects in understanding reaction dynamics involving weakly bound nuclei has been the measurement difficulties in disentangling different reaction processes. Exclusive measurements are essential to unfold those reaction channels~\cite{Maso92,Meij85}. The breakup events from low lying states of ejectile lead to very small relative angles between the breakup fragments. The cross sections of these reaction channels are also very small compared to the elastic scattering cross sections around the grazing angle. These difficulties are overcome by using highly segmented large area Si-detectors~\cite{Cant15,Pand16,Chat16,Luon11,Rafi10,Papk07}. A number of silicon detector arrays such as GLORIA~\cite{Gloria}, MUST2~\cite{Must1} and TIARA~\cite{Tiara} have been built for the study of nuclear reactions. 

Along with the high precision and complex detection setup, the accurate calibration and detection efficiency are also crucial to deduce the absolute cross section from the measured yields~\cite{Torr13,Rees15,Uroi15,Kaya17,Liu18}. The detection efficiency can be estimated by  the Jacobian coordinate transformation as well as using Monte Carlo simulation~\cite{Fuch82,Toki99}. Coordinate transformations and the resulting transformations of the cross sections using Jacobians was discussed for reactions with three particles in the exit channels in Ref.~\cite{Fuch82}. From an experimental point of view Monte Carlo simulation method is preferable over the Jacobians method to account for detection threshold, energy and angular resolution of the detector, beam emittance, beam energy resolution, and multiple-scattering in a thick target foil etc.~\cite{Toki99}. The equivalence of the two methods within statistical uncertainties was also shown in Ref.~\cite{Toki99}. In the present work, we have developed a Monte Carlo simulation code to estimate the efficiency for coincident detection of the breakup fragments and to interpret the observables of different breakup processes. 

The present paper reports detailed description of a Monte Carlo simulation using 3-body kinematics. The method for extraction of absolute cross section and reduction of the various systematic uncertainties are described. The paper is organized as follows: the details of the measurement and analysis of the data are discussed in Sec.~\ref{DataReduction}. The details of the Monte Carlo simulation is given in the Sec.\ \ref{MonteCarloSimulation}. The method to extract absolute cross section is presented in Sec.~\ref{NormalizationConstantSetup}. The identification of different reaction channels from measured coincidence data and simulation are discussed in Sec.\ \ref{MonteCarloEff}-\ref{AsymmetryEnergySpectra}. A summary of the present work along with the applications are discussed in Sec.\ \ref{sum}.

\begin{figure}
\begin{center}
\includegraphics[trim = 15mm 20mm 15mm 160mm, clip,width=80mm,angle =0]{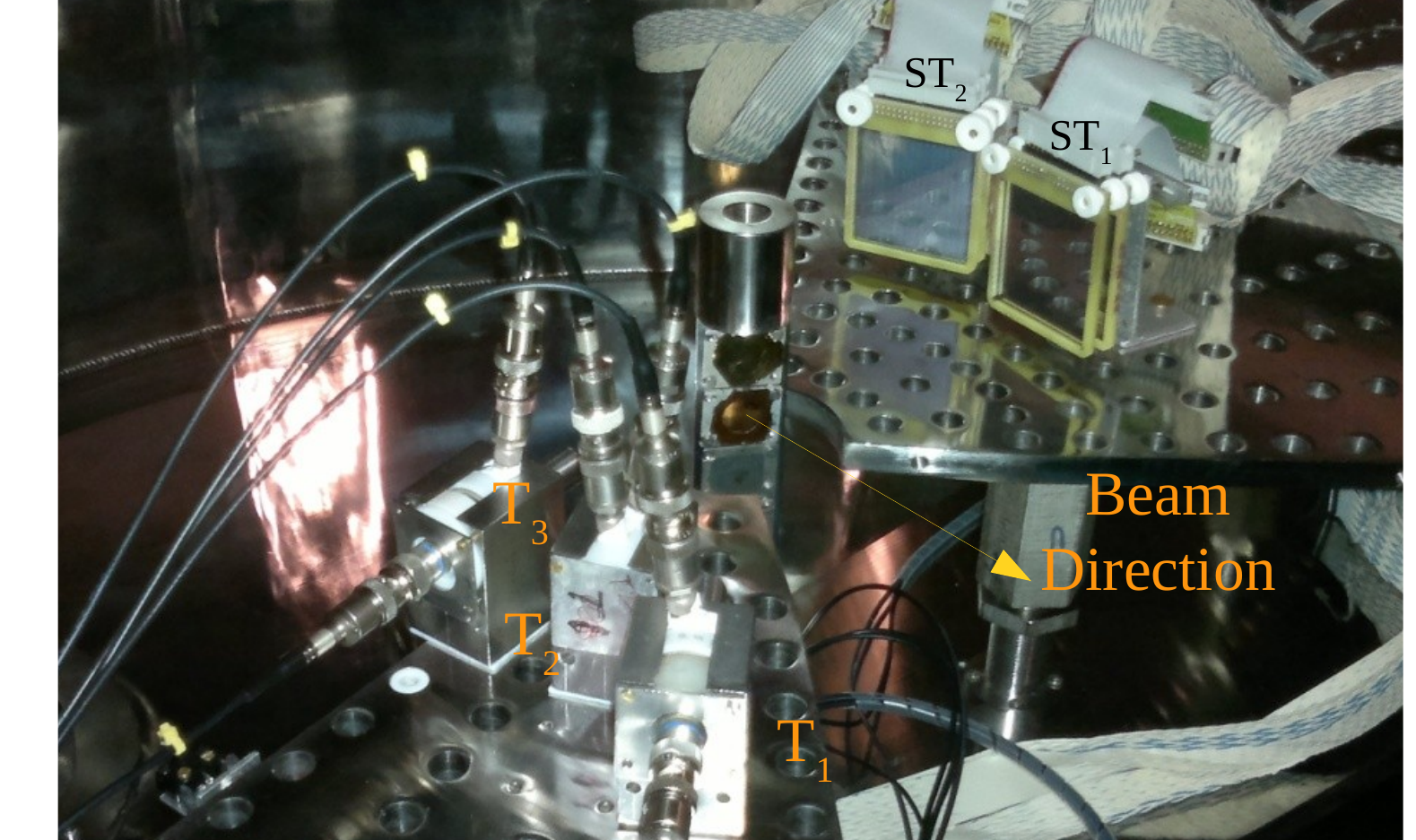}
\caption{Picture of the experimental setup: two telescopes consisting segmented Si-detectors and three telescopes consisting discrete Si surface barrier detectors are shown.}
\label{setup}
\end{center}
\end{figure}  
\section{Experimental details and data reduction}\label{DataReduction}
The experiment was performed at the Pelletron-Linac Facility at TIFR, Mumbai, India. The detectors were mounted
inside a 1.5 m diameter scattering chamber having PLC controlled rotatable arms and target ladder with adjustable height as well as angle of orientation. Self-supporting $^{93}$Nb (thickness $\sim 1.75$ mg/cm$^2$) and Bi (thickness $\sim 0.5$ mg/cm$^2$) foils were used as target. Two telescopes consisting of segmented large area Si-detectors were used. The active area and width of each strip of those segmented detectors were 50.0 x 50.0 mm$^2$ and 3.1 mm, respectively. Thicknesses of $\Delta E$ and $E$ detectors were 50 $\mu$m and 1.5 mm, respectively. The $\Delta E$ detectors were single-sided and the $E$ detectors were double-sided with 16 strips having 256 pixels. The typical separation between the $\Delta E$ and $E$ detectors was $\sim$7 mm. Both the telescopes were mounted at a distance of 16 cm from the target on a movable arm inside the scattering chamber. In this geometry, the cone angle between the two detected fragments ranged from $1^\circ$ to $24^\circ$. Three Si surface-barrier detector telescopes (thicknesses: $\Delta E$ $\sim$ 20-50 $\mu$m, $E$ $\sim$ 450-1000 $\mu$m) were also used. The main purpose of those telescopes is to measure the elastic scattering angular distribution at forward angles ($20^\circ$-$30^\circ$) where the count rate is too high for the strip detectors to cope with and to do comparative study of measured inclusive cross-sections obtained from segmented detectors. Two Si surface-barrier detectors (thickness $\sim 300$ $\mu$m) were kept at $\pm20^\circ$ for absolute normalization. The effect due to beam wandering in the horizontal plane was minimized by taking the geometrical mean of the yields of the two monitor detectors. The data were collected in an event by event mode, with the trigger generated from $E$ detectors. The detectors were calibrated using the known $\alpha$ energies from a $^{239}$Pu-$^{241}$Am-source and the $^7$Li~+~$^{12}$C reaction at 24 MeV~\cite{Park07}.

\begin{figure}
\begin{center}
\includegraphics[trim = 45mm 20mm 45mm 23mm, clip,width=80mm,angle =0]{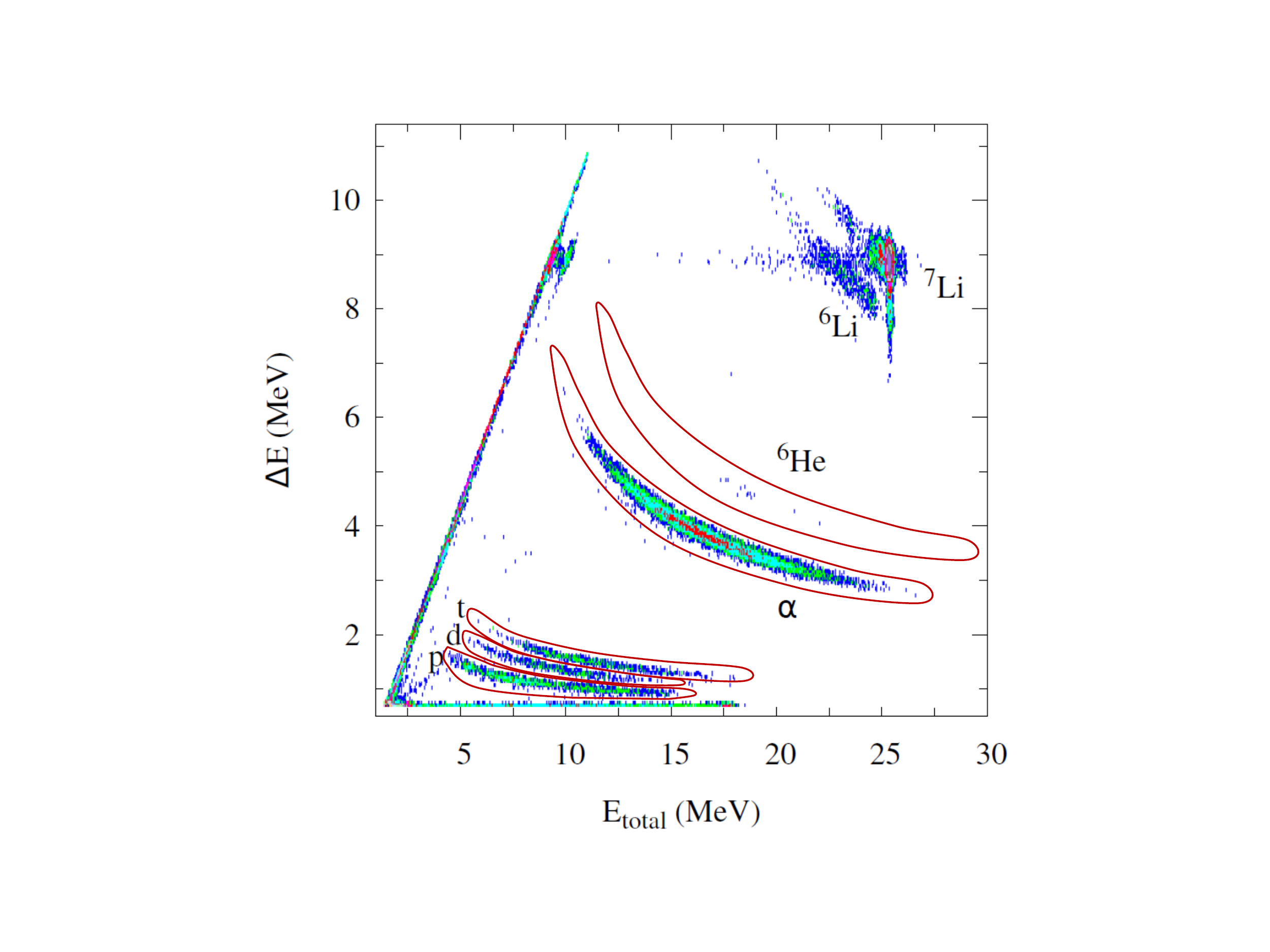}
\caption{Typical particle identification spectrum of $\Delta E$ vs $E_{\rm tot}$ for $^7$Li + $^{93}$Nb reaction at energy E$_{\rm lab}$=27.7 MeV and $\theta_{\rm lab}$~=~60$^{\rm o}$. The banana shaped gates are also marked for identification of different particles.}
\label{delE_Etotal}
\end{center}
\end{figure}

The analysis of the list mode data were carried out using the method discussed below. Particles were identified using the energy loss information from the $\Delta E$ and $E$ detectors. A typical two dimensional spectrum of $\Delta E$ vs $E_{\rm tot}$ for $^7$Li + $^{93}$Nb system at energy E$_{\rm lab}$=27.7 MeV and $\theta_{\rm lab}$~=~60$^{\rm o}$ is shown in Fig.~\ref{delE_Etotal}. A good charge and mass resolution has been achieved which allowed the separation of all the isotopes of Z = 1, 2, and 3 nuclei. From the two dimensional $\Delta E-E$ spectra different banana shaped gates were created corresponding to the identified particles $p, d, t, \alpha,$ and $^6$He as shown in the Fig.~\ref{delE_Etotal}. The use of those banana shaped gates are discussed later.

\begin{figure}
\begin{center}
{\includegraphics[trim = 0mm 0mm 3mm 0mm, clip,width=80mm,angle =0]{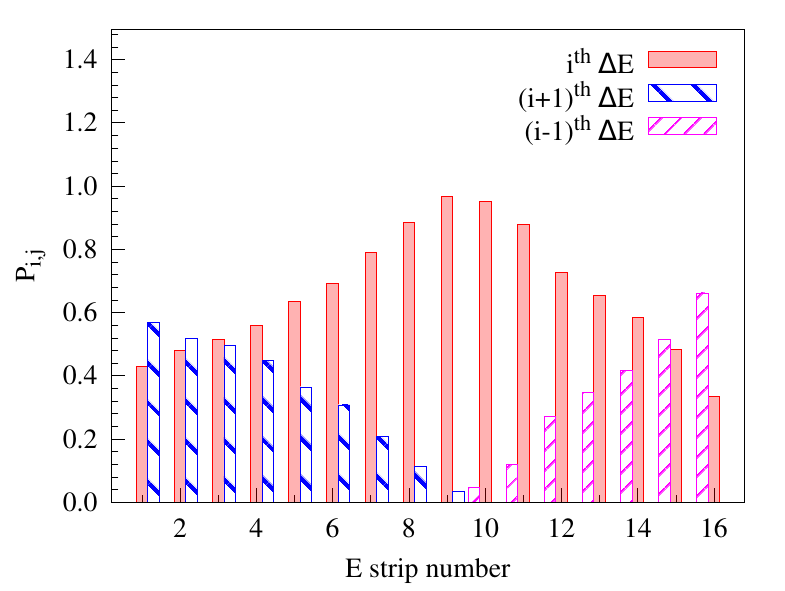}
\put (-185,156) {\small (a)}}
%\vskip -15mm
\includegraphics[trim = 0mm 20mm 0mm 0mm, clip,width=80mm,angle =0]{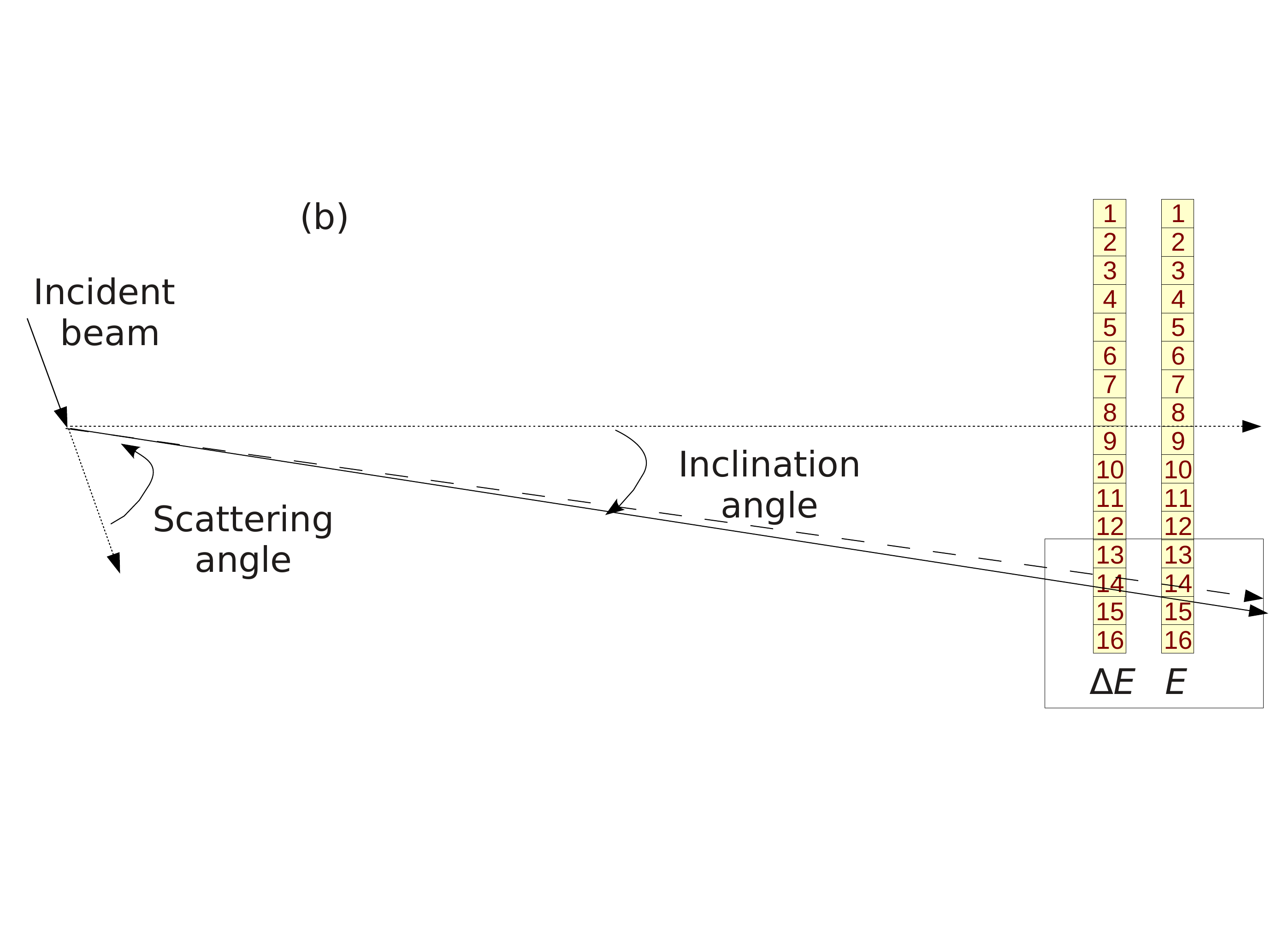}
\vskip -15mm
\includegraphics[trim = 230mm 20mm 0mm 118.5mm, clip,width=30mm,angle =0]{fig3b.pdf}
\caption{(a) The measured probabilities $P_{i,j}$ of elastically scattered particles in different strips of $E$ detector and corresponding $\Delta E$ detector for the $^7$Li+$^{209}$Bi reaction at E$_{\rm lab}$ = 27.9 MeV. (b) Schematic view of the telescope consisting of $\Delta E$ and $E$ strip detectors along with the inclination angle dependency on the hit pattern in strips of $\Delta E$ detector and corresponding $E$ detector. The zoomed view of the rectangular region is shown in the inset.}
\label{inclination}
\end{center}
\end{figure} 

In Fig.~\ref{inclination}(a), the probability of a particle hitting the $i^{th}$ vertical strip of the $E$ detector and $j^{th}$ vertical strip of the $\Delta E$ detector are shown for the elastic scattering of $^7$Li from $^{209}$Bi target at energy E$_{\rm lab}$ = 27.9 MeV. The indexes $i$ and $j$ run from 1 to 16 and correspond to vertical strip number of $E$ and $\Delta E$ detectors respectively. With reasonably aligned $\Delta E$ and $E$ detectors of the current setup, $j$ = $(i-1)$ to $(i+1)$ have been considered for each $i$, as the probability of hitting other strips with $|i-j|>1$ is found to be negligibly small. The probability is defined as $P_{i,j}={\frac{{\rm Y}(E^i,{\Delta}  E^j)}{{\rm Y}(E^i)}}$, where, Y($E^i, {\Delta}  E^j)$ is the yield of the elastically scattered particle corresponding to $i^{\rm th}$ strip of $E$ and $j^{\rm th}$ strip of the $\Delta E$ detector and ${\rm Y}(E^i)$=$\sum _j$Y($E^i,{\Delta} E^j$) is the total elastic yield obtained in the $i^{\rm th}$ strip of the $E$ detector. Although, the probability $P_{i,i}$ is close to 1 for the central region of the detector, the probability $P_{i,j}$ (i$\neq$j) corresponding to the detection in the adjacent strip increases as going from the center to the edge of the detector. This is mostly due to the flat shape of the detectors and can be understood from the schematic given in Fig.~\ref{inclination}(b). In the figure, the particle incident on the detector are shown for two different inclination angle by dashed and solid lines. Although, for both the trajectory the particle hit the 14$^{th}$ strip of the $\Delta E$ detector, depending on the inclination 14$^{th}$ or 15$^{th}$ strip of the $E$ detector may get fired.

The observed hit pattern in the strips shown in Fig.~\ref{inclination}(a) was used to optimise  in searching of the coincidence events in the offline data analysis. A searching algorithm is developed for the coincidence data reduction. In that algorithm, after getting a proper energy signal in the i$^{th}$ strip of the $E$ detector, search for the energy signal in the $\Delta E$ detector were carried out considering the hit pattern shown in Fig.~\ref{inclination}(a).  The correct combination of the set of energy loss information $\Delta E$ and $E$ were checked from the particle identification conditions using the banana shaped particle identification gates discussed above. Identification became complicated when the coincident pair hit the same strip of a given detector. In case of double sided detectors, these events could have been recovered by matching their energy to the sum of the signals measured on the other side of the detector. Taking care of those events is supposed to improve the statistics and to be important for the events corresponding to small relative angles between the breakup fragments. The present detector setup consists of single sided $\Delta E$ detectors, hence, we have rejected those events and same was taken care in the simulation to extract absolute cross sections.

Detected particles were tagged by kinetic energy ($E$), identity (\textit{A, Z}) and scattering angle ($\theta$, $\phi$) with respect to the beam axis.  The relative angles ($\theta_{\rm rel}$) between the fragments were calculated from the measured scattering angles ($\theta_1$, $\phi_1$; $\theta_2$, $\phi_2$) using the relation 

\begin{equation}\label{thetarel} 
\theta_{\rm rel}=cos(\theta_1)cos(\theta_2)+sin(\theta_1)sin(\theta_2)cos(\phi_1-\phi_2),
\end{equation} 

The fragments' mass, kinetic energy ($E_1, E_2$) and  $\theta_{\rm rel}$  were used to calculate their relative energy ($E_{\rm rel}$) from the relation
\begin{equation} 
E_{\rm rel}=\frac{m_2E_1+m_1E_2-2(m_1E_1m_2E_2)^{1/2}cos(\theta_{\rm rel})}{m_1+m_2},
\end{equation} 

The excitation energy of the ejectile prior to breakup was obtained by adding the breakup threshold to the measured $E_{\rm rel}$. The excitation energy of  the target-like  nuclei was determined using the missing energy technique.

\section{Monte Carlo simulation}\label{MonteCarloSimulation}
\begin{figure}
\begin{center}
\includegraphics[trim = 22mm 26mm 15mm 5mm, clip,width=70mm,angle =0]{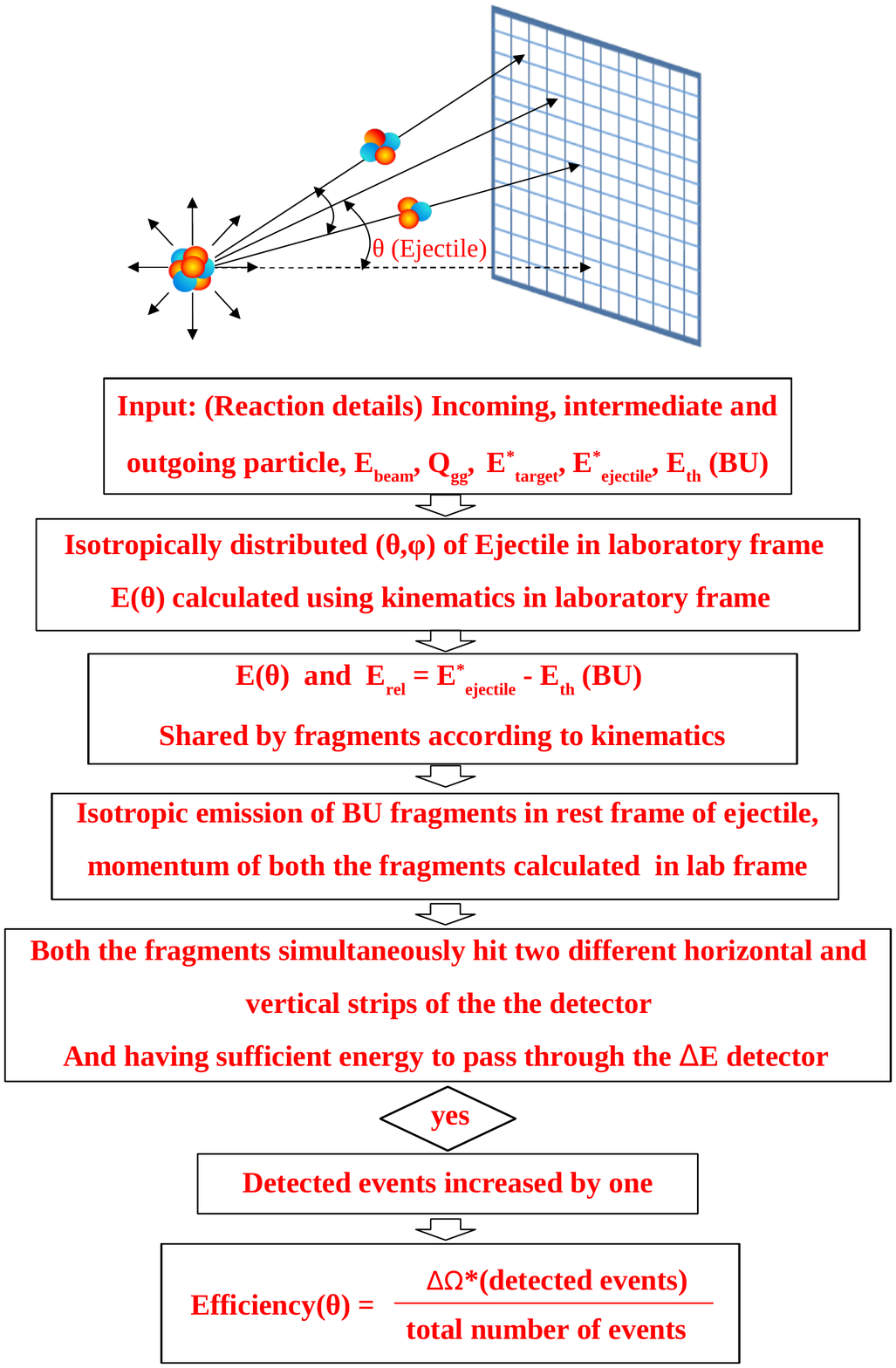}
\caption{The flow chart for the simulation of 3-body kinematics using Monte Carlo technique. Here, E$_{\rm th}$(BU) and Q$_{\rm gg}$ correspond to breakup (BU) threshold and ground state Q-value of the reaction.}
\label{simulationflowchart}
\end{center}
\end{figure}

A Monte Carlo simulation code has been developed using 3-body kinematics to estimate the efficiency of the detection setup consisting of large area segmented Si-detectors. The algorithm of the simulation is presented in the flow chart shown in Fig.~\ref{simulationflowchart}. In the simulation, the scattering angle of the ejectile prior to breakup is generated by an isotropic distribution in a spherical coordinate system. The scattering angle $\theta$ and $\phi$ in lab frame are generated by taking the cosine of a random number having values in between -1.0 to 1.0, and a random number having values in between 0.0 to 2$\pi$, respectively. To achieve a better time efficiency in running the code, instead of throwing the events in 4$\pi$ solid angle, the scattering angle $\theta$ and $\phi$ are restricted to a minimum and maximum values such that the solid angle coverage~$\Delta \Omega$ limited to the region covered in the experiment. The scattered energy of the ejectile is calculated using kinematics taking into account the \textit{Q}-value of the reaction and the excitation energies of the target (E$^*_{\rm target}$) as well as the ejectile (E$^*_{\rm ejectile}$). The coincidence detection efficiency depends on the velocity of the ejectile prior to breakup as well as the relative velocity of the fragments~\cite{Maso92}. The breakup fragment emission in the rest frame of the ejectile is also considered to be isotropic. The velocities of each fragment in the rest frame of the ejectile are calculated using energy and momentum conservation laws. These velocities are added to the velocity of the ejectile prior to breakup to get their velocities in the laboratory frame. Checking of the events to be registered as `the detected event' is carried out in two steps. In the first step, it is checked whether both the breakup fragments hit two different vertical strips of the $\Delta E$-detector and the residual energy after passing through the $\Delta E$ detector is more than the detection threshold of the $E$-detector. In the second step, it is checked whether the fragments hit two different vertical and horizontal strips of $E$ detector. Events satisfying these conditions are considered as detectable events for estimation of the efficiency for coincident detection of the breakup fragments. In the code, the exclusion of the non working strips is also implemented. Misalignment in $\Delta E$ and $E$  detector is also taken into account. The conversion of the energy and scattering angle from the laboratory frame to the rest frame of the target-projectile in event-by-event mode takes care of the Jacobians of the transformation. 
   
\begin{figure}
\begin{center}
\includegraphics[trim = 0mm 0mm 0mm 0mm, clip,width=80mm,angle =0]{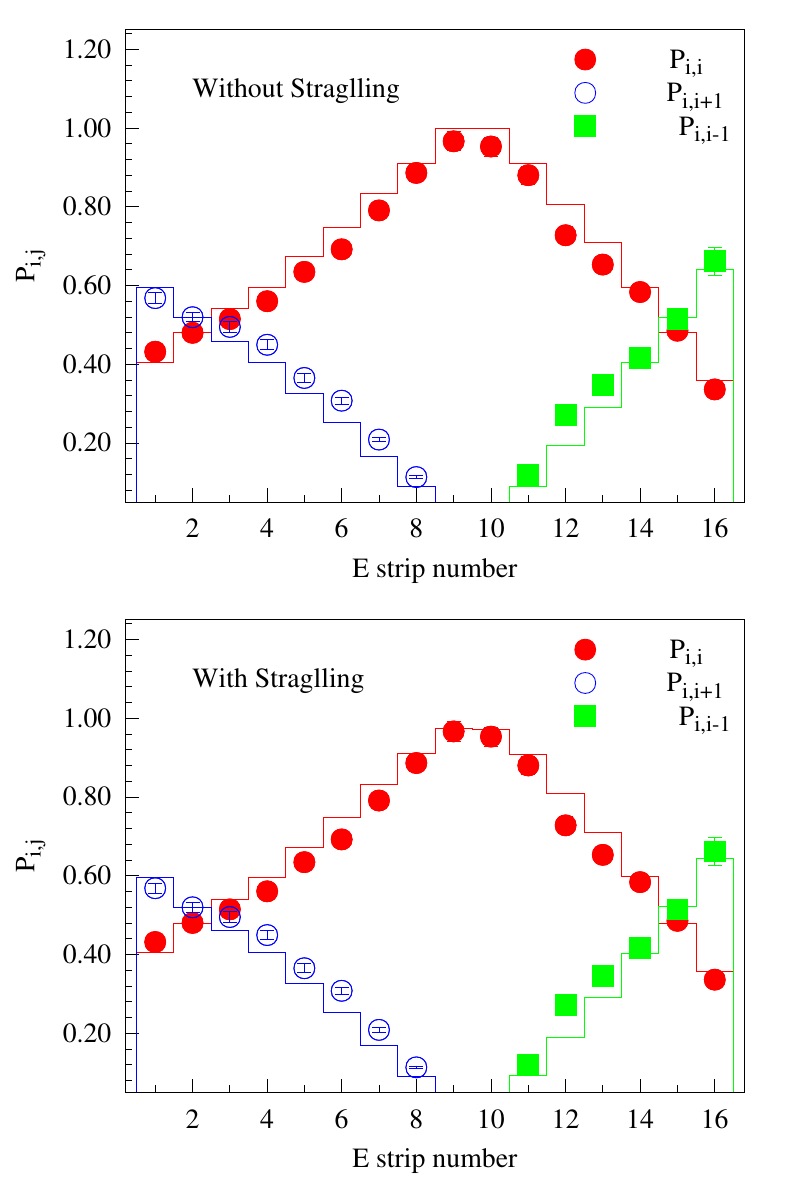}
\put (-185,316) {\small (a)}
\put (-185,144) {\small (b)}
%\vskip -15mm
\caption{The experimental hit distributions $P_{i,j}$ of the elastic scattering for the $^7$Li+$^{209}$Bi system at E$_{\rm lab}$ = 27.9 MeV are compared with simulation (see text for details). The red filled circles, open blue circles, and green filled squares are measured probabilities $P_{i,i}$, $P_{i,i+1}$, and $P_{i,i-1}$, respectively. The simulated results are shown by respective histograms. (a) and (b) are corresponding to the simulated results with and with out consideration of the effect of the straggling.}
\label{simulated_ratio}
\end{center}
\end{figure}

The probability $P_{i,j}$, discussed earlier, is interpreted from simulation. It is found that the position of the detector (effective centre of the detector) normal to the ineraction point in the target and gap between the $\Delta E$ and $E$ detectors are the sensitive parameter for the reproduction of the measured value of $P_{i,j}$. The comparison of the simulation with the data is shown in Fig.~\ref{simulated_ratio}. The method is good to find out the effective centre of the detector and the gap between the $\Delta E$ and $E$ detectors. The effective centre of the detector and gap between the $\Delta E$ and $E$ detectors are important parameters for the estimation of the response of the detection setup accurately. Along with the geometrical inclination, the straggling in $\Delta E$ detector could also be another reason of adjacent strip firing. However, from the estimation of the energy-loss and ranges straggling using {\small LISE} code~\cite{Lise}, it is found that the effect of straggling is very small. The simulated results after consideration of the straggling effect is presented in Fig.~\ref{simulated_ratio}(b). 

It is also noticed that the probability of hitting to the adjacent strips can be reduced by minimizing the gap between $\Delta E$ and $E$ detectors. This observation of large fraction of the hit in the adjacent strips in the $\Delta E$ detector suggests that, inclusion of the events corresponding to adjacent strips are necessary for the gain in statistics.

The estimated detection efficiency of different coincidence events varies due to kinematic focusing, which is found to depend on relative energy of the breakup fragments, energy of the ejectile prior to breakup, mass asymmetry of the breakup fragments, detection threshold, and geometric solid angle of the detection setup. Since energy of the ejectile prior to breakup is decided by the incident beam energy of the projectile, ground state $Q$-value ($Q_{\rm gg}$) of the reaction, the excitation of the ejectile as well as target like nuclei, detection efficiency is also affected by these parameters.

This code is also used to get the efficiency for singles measurement by applying the conditions of no breakup and limiting the detection condition for a single fragment with respective detection thresholds. The estimated efficiency has been verified with the geometric solid angle calculated using the analytic formula. 

\subsection{Determination of normalization constant of the detection setup}\label{NormalizationConstantSetup}
The cross section of a selected reaction channel can be derived as
\begin{equation}\label{finalCrossSection}
\frac{d\sigma}{d\Omega}(\theta)=(\frac{dY(\theta)}{d\theta}/\frac{d\Omega(\theta)}{d\theta})\times \frac{1}{Y_m (\theta _m)} \times (\frac{Z_{\rm P}Z_{\rm T}}{E})^2 \times \frac{1}{K} 
\end{equation}  
where, $\frac{dY(\theta)}{d\theta}$ is the measured yield distribution of the corresponding reaction channel. The  $\frac{d\Omega(\theta)}{d\theta}$ is the detection efficiency of the measurement setup for that selected reaction. $Z_{\rm P}$ and  $Z_{\rm T}$ are the atomic number of the projectile and target, respectively. $E$ is the energy of the incident beam. $Y_m(\theta _m)$ is the measured yield in the monitor detector kept at an angle $\theta _m$. $K=\frac{1}{1.296}\frac{sin^4(\frac{\theta_m}{2})}{d\Omega_m}$ is a constant of the measurement setup. Here, $d\Omega_m$ is the solid angle of the monitor detector. Hence, to extract the absolute cross-sections of the various breakup processes, the normalization constant $K$ is also an important parameter. In case of pure Rutherford scattering, the Eq.~\ref{finalCrossSection} can be simplified and value of $K$ can be estimated as

\begin{equation}
K=\frac{1}{1.296}\frac{dY_{el}(\theta)}{d \Omega} \frac{sin^4(\frac{\theta}{2})}{Y_m(\theta _m)}
\end{equation}

\begin{figure}[h]
\begin{center}
\includegraphics[trim = 0mm 0mm 0mm 0mm, clip,width=80mm]{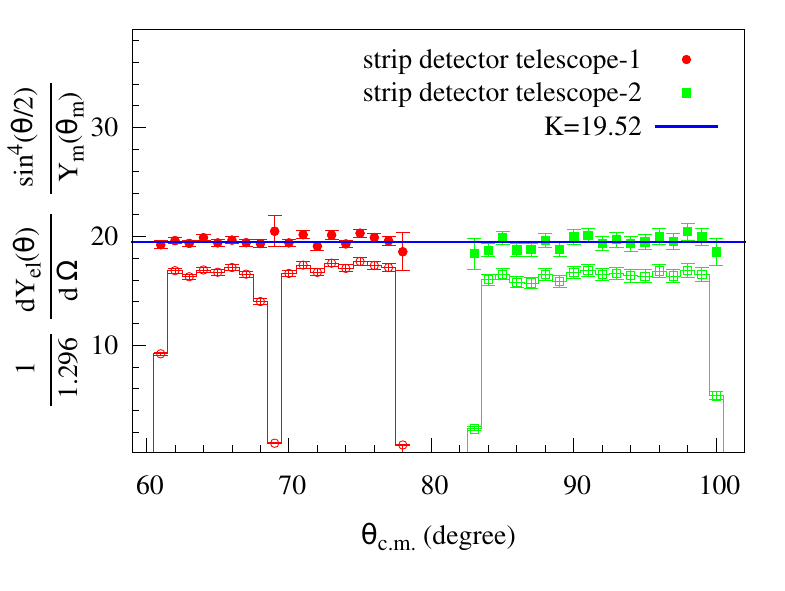}
\caption{\label{figsimulation1} Measured ratio of elastic yields to monitor yields for $^7$Li on $^{209}$Bi at $E_{\rm lab} = 27.9$ MeV are shown by red open circles and green open squares. The extracted $\frac{1}{1.296}\frac{sin^4(\frac{\theta_m}{2})}{d\Omega_m}$ parameter for the strip-detector telescopes 1 and 2 are shown by red solid circles and green solid squares, respectively. The value of $K$ = 19.52 correspond to the fitted value shown by solid line.}
\end{center}
\end{figure}

We have chosen $^7$Li+$^{209}$Bi system, for which Coulomb barrier is $\sim$30 MeV. The measurement of elastic scattering at $E_{\rm lab}$ = 27.9 MeV can be considered as Rutherford scattering in the measured angular range. In Fig.~\ref{figsimulation1}, the normalized yields $Y_{el}(\theta)$ for telescopes 1 and 2 are shown by open circles and squares, respectively. The angular distribution of the normalized yields for telescope 1 is found to be relatively small at angles corresponding to the edge of the detector as well as almost centre of the detector. The drops of counts at the edges are due to the finite gap between the $\Delta E$ and $E$ detectors as discussed in the Sec.\ \ref{MonteCarloSimulation}. The dip in the angular distribution at the angles 68$^o$ and 69$^o$ is due to the absence of the electronic channel corresponding to 7$^{th}$ strip of the telescope 1. The efficiency corrected normalized yields $\frac{1}{1.296}\frac{dY_{el}(\theta)}{d \Omega}\frac{sin^4(\frac{\theta}{2})}{Y_m(\theta _m)}$ is plotted in Fig.~\ref{figsimulation1}. The efficiency correction leads to smooth angular distributions. The solid circles and squares in the figure are corresponding to the telescopes 1 and 2, respectively. The $\chi^2$ minimized fitted value is found to be 19.52. The experimentally determined the value of normalization constant $K$ for the measurement setup is found to be close to the calculated value 20.1. The method of analysis for extracting absolute cross section discussed above minimizes the uncertainties due to the target thickness, beam current and solid angle of the detection setup. 

\subsection{Interpretation of the measured observables}\label{MonteCarloEff}
\begin{figure}[h]
\begin{center}
\includegraphics[width=80mm]{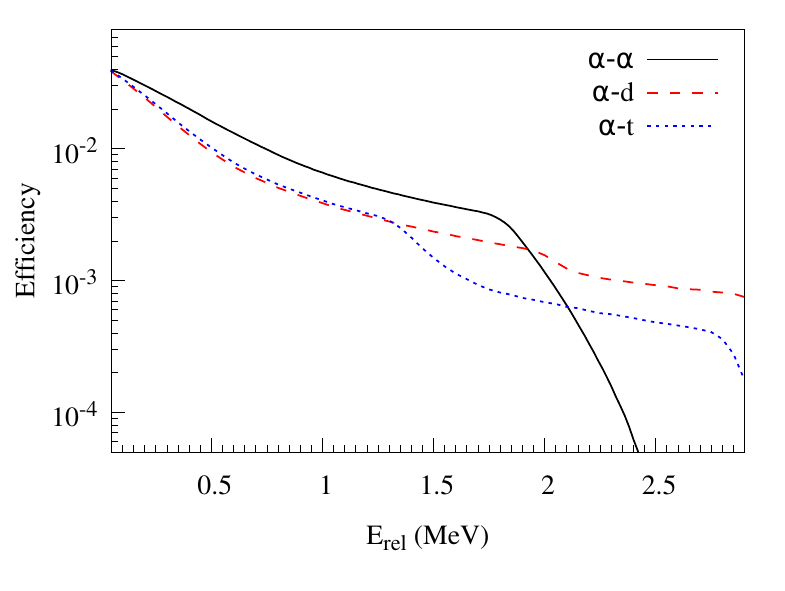}
\caption{\label{Eff_Erel} The estimated detection efficiency of a Si-strip detector telescope kept at a distance of $\sim$15 cm. The efficiencies corresponding to  $\alpha$-$\alpha$, $\alpha$-$d$ and $\alpha$-$t$ events as a function of relative energy are shown.}
%\vskip -5mm
\end{center}
\end{figure}

The simulation has also been used to interpret different measured observables, e.g. energy-angle correlations and energy spectra of the outgoing fragments, etc. Three types of coincidence events, $\alpha$-$\alpha$, $\alpha$-$d$ and $\alpha$-$t$ have been studied for $^7$Li+$^{93}$Nb reaction~\cite{Pand16}. To explain the $\alpha$-$\alpha$ and $\alpha$-$d$ coincidence events, two step processes 1$p$-pickup and 1$n$-stripping reaction mechanisms followed by breakup of $^8$Be and $^6$Li have been considered, respectively, whereas the $\alpha$-$t$ events have been reproduced from the direct breakup of $^7$Li due to inelastic excitation to the continuum and resonance states. The ground state $Q$-value ($Q_{\rm gg}$) and the breakup threshold ($E_{\rm th}$) for all the three coincidence events are reported in Table~\ref{Qgg_BUthreshold}. The efficiency corresponding to these events are shown in Fig.~\ref{Eff_Erel} as a function of the relative energy of the breakup fragments. The sharp fall of the efficiency at large value of the relative energy is due to the insufficient energy of any one of the breakup fragments to pass through the $\Delta E$ detector or below the threshold of the $E$ detector.

\begin{table*}[h]
\begin{center}
\caption{\label{Qgg_BUthreshold}Ground state $Q$-value $Q_{\rm gg}$ and breakup threshold $E^{\rm BU}_{\rm th}$ of the different coincident events.}
%\vskip 5 mm
%\resizebox{\textwidth}{!}{
\begin{tabular}{|c|c|c|c|c|}
\hline
Coincident & \multicolumn{2}{ |c| }{Reaction} &$Q_{gg}$ &$E^{\rm BU}_{\rm th}$\\ 
 events & \multicolumn{2}{ |c| }{mechanism} &(MeV) &(MeV)\\ \hline
{$\alpha$-$\alpha$} & {1$p$-pickup} & $^{93}$Nb($^7$Li,$^8$Be$^*$) & 11.21 & {0.0}\\ \hline
{$\alpha$-$d$} &  {1$n$-stripping} & $^{93}$Nb($^7$Li,$^6$Li$^*$) & -0.02 & {1.47}\\ \hline
 {$\alpha$-$t$} &  {inelastic } & $^{93}$Nb($^7$Li,$^7$Li$^*$) & 0.0 & {2.47}\\ \hline
\end{tabular}
%}
\end{center}
\end{table*}

\begin{figure}
\begin{center}
\includegraphics[trim = 0mm 0mm 0mm 0mm, clip,width=70mm,angle =0]{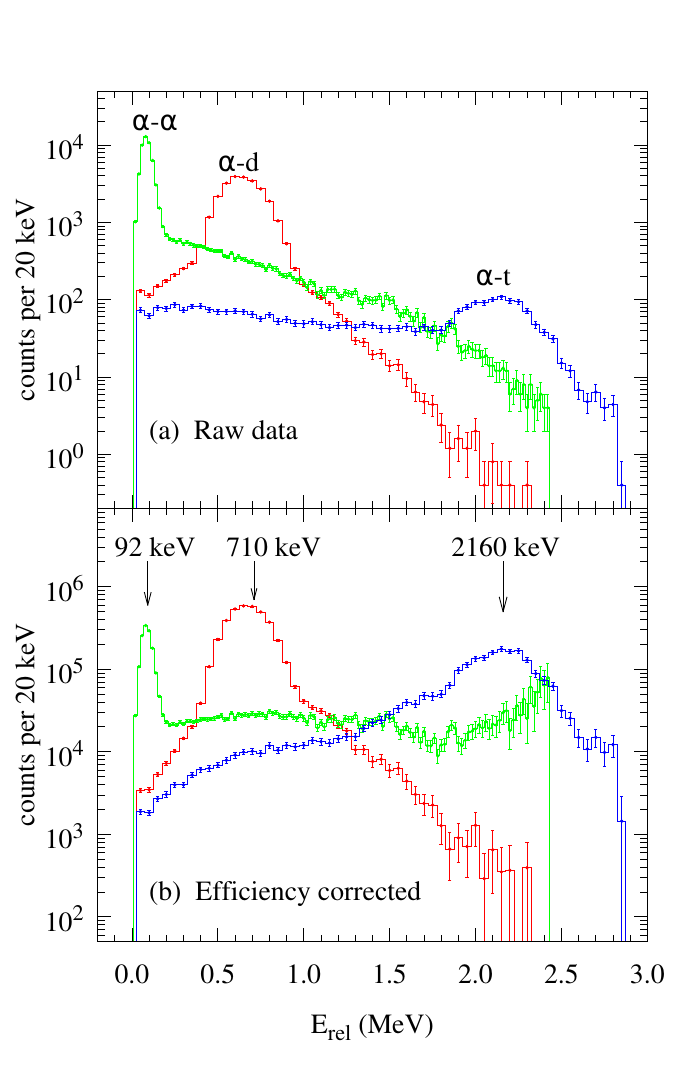}
\caption{(a) The extracted relative energy spectra, (b) efficiency corrected relative energy spectra of $\alpha$-$\alpha$, $\alpha$-\textit{d} and $\alpha$-\textit{t} for the $^7$Li+$^{93}$Nb system at $E_{\rm lab}$ = 27.7 MeV and $\theta_{\rm lab}$ = 60$^\circ$.}
\label{ErelNb28}
\end{center}
\end{figure}

The measured relative energy spectra before and after the efficiency correction for $\alpha$-$\alpha$, $\alpha$-$d$ and $\alpha$-$t$ events are shown in~Fig.\ \ref{ErelNb28}. It is clear that efficiency corrections modifies the spectra shape significantly. The shapes of the peaks at 0.71 MeV and 2.16 MeV in the relative energy spectra of $\alpha$-$d$ and $\alpha$-$t$ are improved. The two peaks are corresponding to the breakup of $^6$Li and $^7$Li from 3$^+$ and $\frac{7}{2}^-$ states, respectively. The rising trend in the $\alpha$-$\alpha$ spectrum peak above the $E_{\rm rel}$ = 2.0 MeV could be part of the 2$^+$ ($E^*$ = 3.0 MeV) state~\cite{Till05}. These observations show the importance of the efficiency estimation for proper identification of different reaction channels. 

\begin{figure}
\begin{center}
\includegraphics[width=80mm]{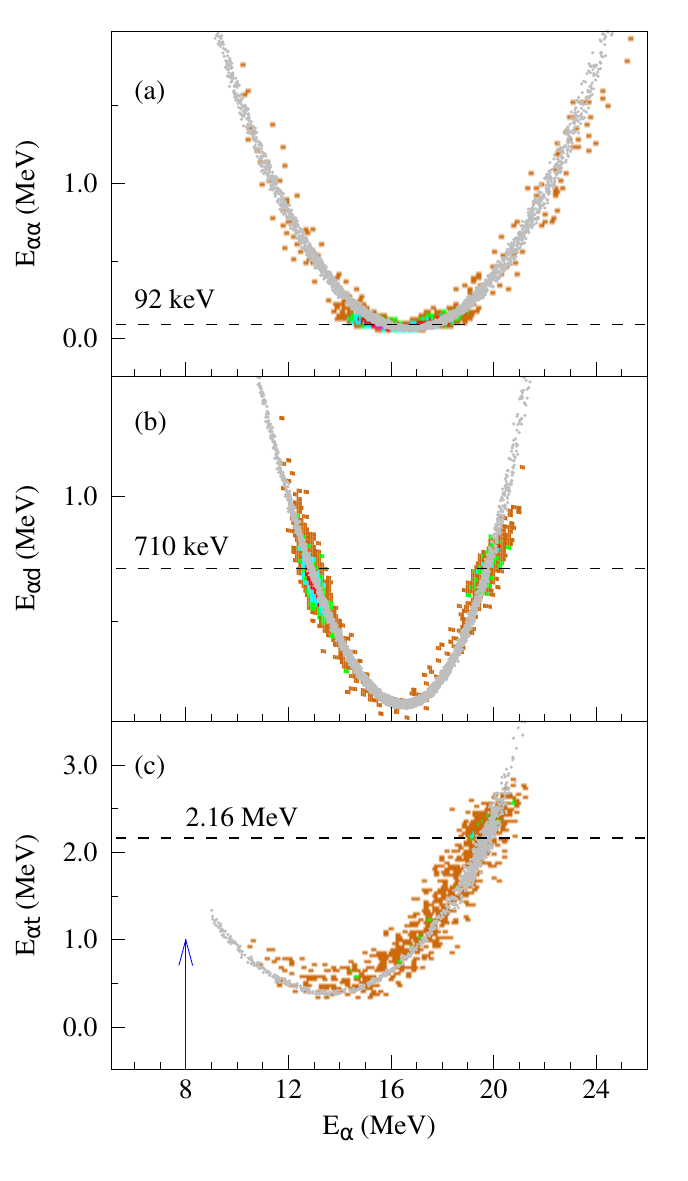}
\caption{\label{EaErelSimulation4thesis} Simulated energy correlation spectra for $^7$Li + $^{93}$Nb reaction at $E_{\rm lab} = 27.7$ MeV are compared with the measured data. Grey colored points are obtained from simulation and colored points are corresponding to the data. (a) $E_{\alpha}$ vs. $E_{\alpha \alpha}$($^8$Be$\rightarrow \alpha+\alpha$) corresponding  to  $\theta_{\rm rel}^{\alpha \alpha}$ = $3^\circ$. (b) $E_{\alpha}$ vs. $E_{\alpha d}$($^6$Li$\rightarrow \alpha+d$) corresponding to $\theta_{\rm rel}^{\alpha d}$ = $10^\circ$. (c) $E_{\alpha}$ vs. $E_{\alpha t}$($^7$Li$\rightarrow \alpha+t$)  corresponding  to  $\theta_{\rm rel}^{\alpha t}$ = $15^\circ$.  The arrow on the x axis indicates the detection threshold.}
%\vskip -5mm
\end{center}
\end{figure}

\subsection{Correlated energy spectra}\label{CorrelatedEnergySpectra}
The measured energies of the breakup fragments corresponding to the different states of the ejectile prior to breakup have been reproduced using the simulation. The measured energy correlations of breakup fragments, correlation of relative energy and energy of individual fragment, and energy angle correlations  have been reproduced considering different excited states of the target like nuclei and other parameters discussed earlier. The comparison of simulated and the measured energy correlation spectra for various breakup processes are shown in Fig.~\ref{EaErelSimulation4thesis}. The relative energy between the breakup fragments are plotted as a function of energy of the outgoing $\alpha$-particle, for the the breakup of $^8$Be, $^7$Li and $^6$Li  in Fig.~\ref{EaErelSimulation4thesis}(a), (b) and (c), respectively. For all the three cases the correlation  follows a parabolic trend. 

\begin{figure}
\begin{center}
{\includegraphics[trim = 100mm 70mm 60mm 80mm, clip, width=80mm]{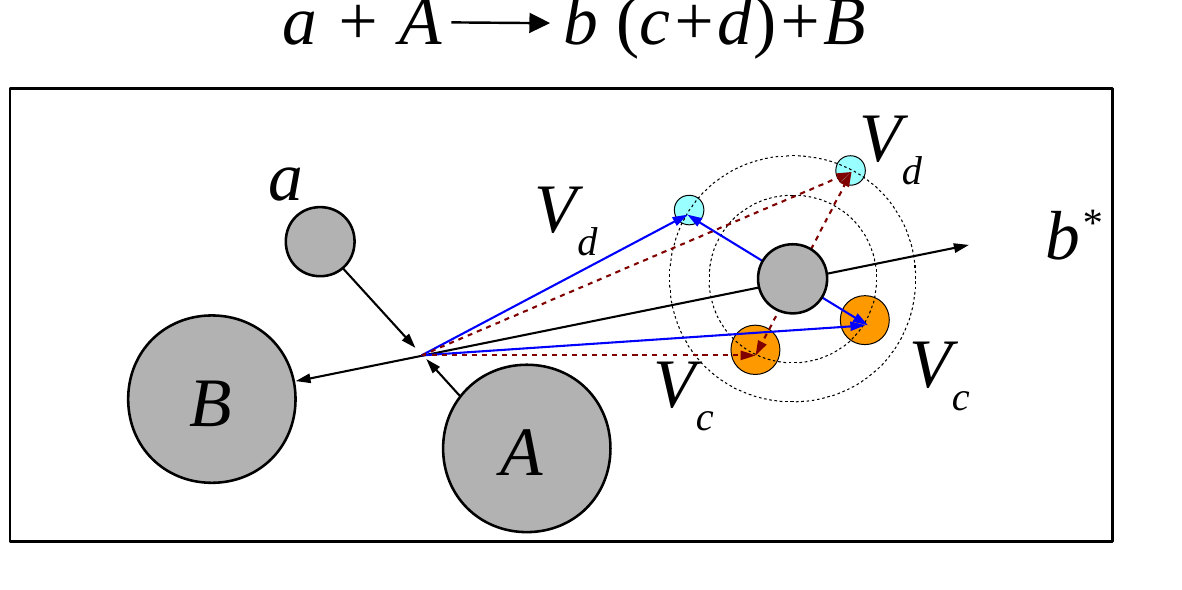}}
%\vskip -5mm
\caption{\label{kinematics} Schematic of the breakup kinematics of a nucleus $b$ into two fragments $c$ and $d$ in A(a,b)B reaction. The velocities of the breakup fragments are shown for two specific breakup orientation by dashed and solid lines. The velocity of the heavier fragment $c$ (lighter fragment $d$) is more than the other for the breakup orientation shown by solid (dashed) line.}
%\vskip -5mm
\end{center}
\end{figure}

\begin{figure}
\begin{center}
\includegraphics[trim = 0mm 0mm 0mm 0mm, clip, width=80mm]{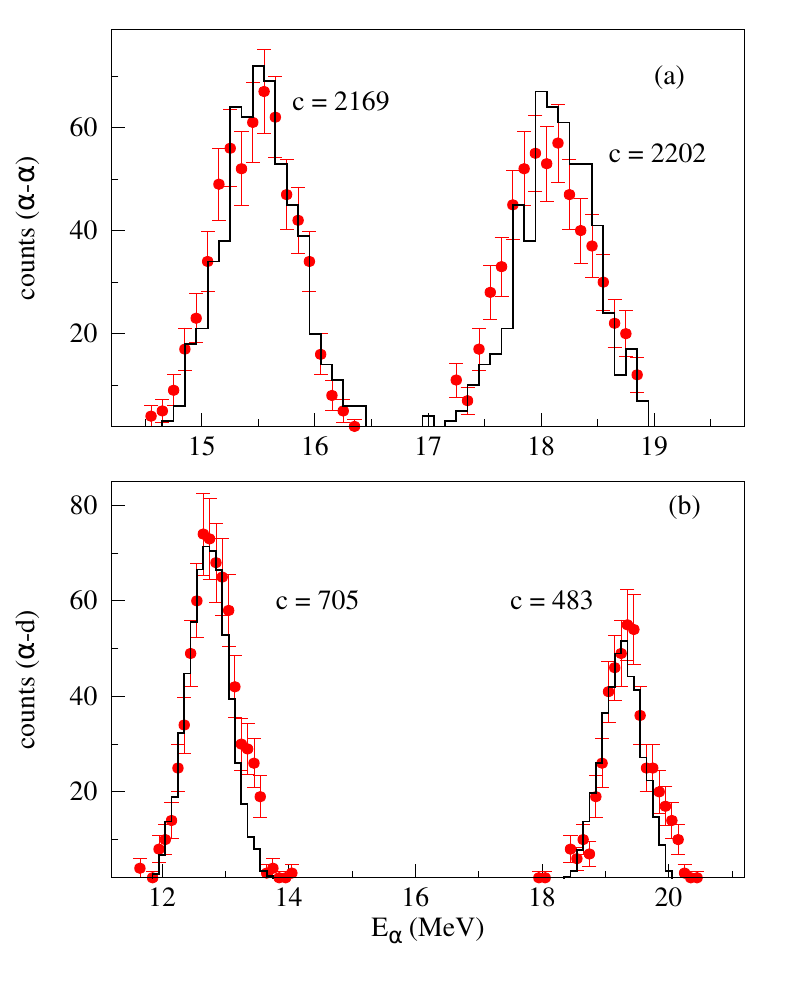}
\caption{\label{aa_ad_at_energy_spectra28} The simulated energy spectra of the $\alpha$-fragment from the breakup of $^8$Be and $^6$Li for $^7$Li+$^{93}$Nb reaction at E$_{\rm lab}$ = 27.7 MeV. The filled circles are the measured data. The number of counts `c' under the individual peaks are also given. The energy spectra of the $\alpha$ particles due to the 1$p$-pickup followed by breakup of $^8$Be$_{\rm g.s.}$ is shown in (a). Here the data is restricted for excitation energy of target nuclei to 3.0 $\leq$ $E^*_{\rm target} \leq$ 4.0 MeV and $\theta_{\rm rel}^{\alpha \alpha}$ = $3^\circ$. (b) corresponds to the 1$n$-stripping followed by breakup of $^6$Li$_{3^+}$ data with $\theta_{\rm rel}^{\alpha d}$ = $10^\circ$ and 0.0 $\leq$ $E^*_{\rm target} \leq$ 0.5 MeV.}
%\vskip -5mm
\end{center}
\end{figure}
\subsection{Asymmetry of yields in correlated energy spectra}\label{AsymmetryEnergySpectra}
The energy of the breakup fragments detected in coincidence depends on the velocity of the ejectile prior to breakup and the velocities of the breakup fragments in the rest frame of the ejectile. A schematic of the velocity diagram for the breakup fragments is shown in~Fig.\ref{kinematics} for the $A$($a,b\rightarrow c +d$)$B$ reaction.     
Two specific breakup orientations are shown in the figure by solid and dashed lines. It can be seen from the figure that the velocity of the fragment $c$ is more than the fragment $d$ for the breakup orientation shown by solid line. However, for the another breakup orientation shown by the dashed line the velocity of the fragment $d$ is more than the fragment $c$. Consequently, two groups with high and low energy of the same fragment are expected in the outgoing channels. In the measurements, two types of coincidence events of high (low) energy $\alpha (d)$ and low (high) energy $\alpha (d)$ have also been observed as shown in Fig.~\ref{EaErelSimulation4thesis}(b) and Fig.~\ref{aa_ad_at_energy_spectra28}(b). The yields corresponding to the high and low energy $\alpha$ particle are found to be asymmetric. Such an asymmetry was also observed in Ref.~\cite{Sant09} and different cross sections for the high and low energy fragments were reported. However, in the present work, the observed asymmetry has been reproduced by the simulation and consequently  consistent cross sections for the high and low energy $\alpha$ particles have been obtained. 

\begin{figure}
\begin{center}
{\includegraphics[trim = 0mm 0mm 0mm 2mm, clip, width=80mm]{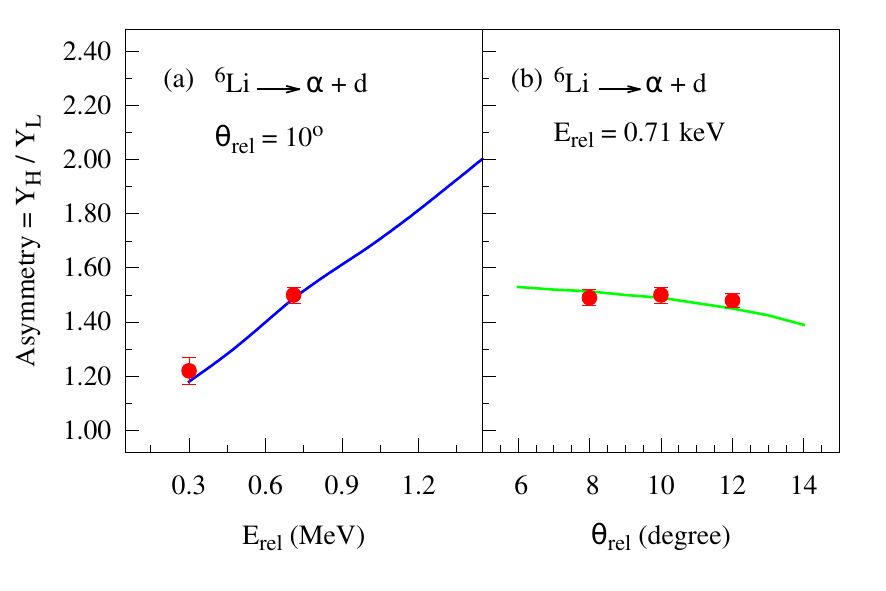}}
%\vskip -5mm
\caption{\label{Asymetry} Measured and simulated asymmetry of $\alpha -d$ events due to the breakup of $^6$Li in the $^{93}$Nb($^7$Li$\rightarrow$ $^6$Li) reaction at $E_{\rm lab}$ = 27.7 MeV. Dependency of asymmetry as a function of (a) $E_{\rm rel}$ and (b) $\theta _{\rm rel}$ are shown.}
%\vskip -5mm
\end{center}
\end{figure}

The simulated energy spectra of breakup fragments $\alpha - \alpha$,  $\alpha - d$, and $\alpha - t$ due to the breakup of $^8$Be, $^6$Li, and $^7$Li, respectively are estimated. The energy spectra of $\alpha$-particle from the breakup of $^8$Be and $^6$Li are shown in Fig.~\ref{aa_ad_at_energy_spectra28}. Although the asymmetry is observed in $\alpha - d$ events, it is  not found in energy spectra of $\alpha - \alpha$ events. The asymmetry arises due to the different kinematic focusing for the coincident fragments. In case of $\alpha - \alpha$ events, the same mass of the two $\alpha$-particle leads to the same kinematic focusing and hence no asymmetry has been observed. It is also found that the asymmetry depends on the difference in mass of the two fragments, relative energy, relative angle and the scattering angle of the ejectile prior to breakup. The asymmetry is defined as the ratio of $Y_H$ and $Y_L$. Where, $Y_H$ and $Y_L$ are the yields of the high energy (forward going) and low energy (backward going) $\alpha$-fragments. The dependency of asymmetry as a function $E_{\rm rel}$ and $\theta _{\rm rel}$ are shown in Fig.~\ref{Asymetry}(a) and (b), respectively for the breakup of $^6$Li in $^{93}$Nb($^7$Li,$^6$Li$\rightarrow \alpha +d$) reaction at $E_{\rm lab}$ = 28 MeV. The simulated results are found to be in good agreement with the measured values.

%The validity of the simulation has also been checked by reproducing the asymmetry of $\alpha-d$ coincidence events for $^6$Li + $^{209}$Bi reaction of Ref.~\cite{Sant09}.      

\section{Summary and applications}\label{sum}
In summary, we have addressed various points in the detailed analysis method of the data, that are acquired using segmented large area Si-detectors, to extract absolute cross sections. The details of a 3-body simulation code based on Monte Carlo method are reported. The loss of coincidence events due to the hitting of both the fragments in a same segment of the detector are included in the simulation. The detection threshold and misalignment in the $\Delta E$ and $E$ detectors are also considered for the accurate estimation of the efficiency. The method to extract the actual centre of the detectors and the gap between the $E$ and $\Delta E$ detectors are mentioned. The accurate knowledge of these parameters are found to be important for the estimation of the accurate response of the detection setup, which is responsible for the proper angular distribution measurements and consequently extraction of the absolute cross sections of the different reaction processes. The extracted cross sections of various breakup processes for $^7$Li+$^{93}$Nb system using the methods discussed in the present paper were published in our earlier paper \cite{Pand16}. The interpretation of the various observables from coincidence measurements has also been presented. The observed asymmetry in the yields of forward and backward going breakup fragments are understood from the kinematic focusing. The simulation will be useful in the designing of the efficient experimental setup consisting a large number of segmented detectors for the study of nuclear reactions involving more than two fragments in the outgoing channels.

\textbf{Acknowledgments} \\
We thank Professor S. Kailas and Professor A. Chatterjee for useful discussions.
 
\bibliography{7li}% Produces the bibliography via BibTeX.  7lielal
\end{document}